\documentclass[10pt,prc,aps,twocolumn,showpacs,showkeys
,superscriptaddress,groupedaddress,floatfix]{revtex4-1}
\usepackage{bm}
\usepackage{amsmath}
\usepackage{amsfonts}
\usepackage{amssymb}
\usepackage{graphicx}
\usepackage{graphics}
\usepackage[normalem]{ulem}
\usepackage{soul}
\usepackage[dvips]{color}
\usepackage[latin1]{inputenc}
\usepackage[english]{babel}
\usepackage{anysize}
\newcommand{\be}{\begin{equation}}
\newcommand{\ee}{\end{equation}}
\newcommand{\ba}{\begin{eqnarray}}
\newcommand{\ea}{\end{eqnarray}}

\def\ee{\mbox{$\left(e,e^{\prime}\right)$\ }}
\def\eep{\mbox{$\left(e,e^{\prime}p\right)$\ }}

\begin{document}

\author{R. Gonz\'alez-Jim\'enez}
\author{J.A. Caballero}
\affiliation{Departamento de F\'{\i}sica At\'{o}mica, Molecular y Nuclear,
Universidad de Sevilla,
  41080 Sevilla, Spain}
\author{Andrea Meucci}
\author{Carlotta Giusti}
\affiliation{Dipartimento di Fisica,
Universit\`{a} degli Studi di Pavia and \\
INFN,
Sezione di Pavia, via A. Bassi 6, I-27100 Pavia, Italy}
\author{M.B. Barbaro}
\affiliation{Dipartimento di Fisica, Universit\`{a} di Torino and INFN, \\
Sezione di Torino, Via P. Giuria 1, 10125 Torino, Italy}
\author{M.V. Ivanov}
\affiliation{Grupo de F\'{\i}sica Nuclear, Departamento de F\'{\i}sica 
At\'{o}mica, Molecular y Nuclear,
Universidad Complutense de Madrid, CEI Moncloa, 28040 Madrid, Spain}
\affiliation{Institute for Nuclear Research and Nuclear Energy, 
Bulgarian Academy of Sciences, Sofia 1784, Bulgaria}
\author{J.M. Ud\'{\i}as}
\affiliation{Grupo de F\'{\i}sica Nuclear, Departamento de F\'{\i}sica 
At\'{o}mica, Molecular y Nuclear,
Universidad Complutense de Madrid, CEI Moncloa, 28040 Madrid, Spain}

\title{Relativistic description of final-state interactions
in neutral-current neutrino and antineutrino cross sections}

\date{\today}

\begin{abstract}
We evaluate semi-inclusive neutral-current quasielastic differential 
neutrino and antineutrino cross sections
within the framework of  the relativistic impulse approximation.
 The results of the relativistic mean field and of the relativistic Green's
function models are compared. The sensitivity
to the strange-quark content of the nucleon form factor is also discussed.
 The results of the models are compared with the MiniBooNE experimental data for
neutrino scattering. Numerical predictions for flux-averaged antineutrino
scattering cross sections are also presented.
\end{abstract}


\maketitle

\section{Introduction}
\label{intro}

The results on neutrino oscillations published by
different collaborations \cite{superkam,superkam11,icarus13,sno,minos12,t2k11,minibPhysRevLett13,minibPhysRevD12,PhysRevD.86.052009,PhysRevLett.108.131801,PhysRevLett.108.171803,
An:2012bu,PhysRevLett.108.191802,PhysRevD.74.072003,PhysRevD.64.112007} have raised a
large debate over the properties of
neutrinos that could lead to a more complete understanding of neutrino
physics. Because of the interest in oscillation measurements,
various  experimental neutrino-nucleus differential cross sections
have been presented \cite{bnl,miniboone,miniboonenc,miniboone-ant,Nakajima:2010fp,argoneut}
and are planned in the near future~\cite{minibooneweb,minerva,t2k}.
A clear understanding of neutrino-nucleus reactions with a precise determination of
differential cross sections is crucial for  a
proper analysis of the experimental data.

The MiniBooNE Collaboration  has recently reported \cite{miniboonenc}
a measurement of the neutral-current elastic (NCE)
flux-averaged differential neutrino cross
section on CH$_2$ as a function of the four-momentum transferred squared $Q^2$.
The energy region
considered in the MiniBooNE experiments, with average neutrino energy  of $\approx 0.8$ GeV,
requires the use of a relativistic model with an adequate description of  nuclear
dynamics and current operators.  The relativistic Fermi gas
(RFG) model cannot reproduce the data unless calculations are performed with a
value of the axial mass $M_A$ significantly larger
($M_A = 1.39 \pm 0.11$ GeV/$c^2$)  than the world average value from the
deuterium data of $M_A \simeq 1.03$ GeV/$c^2$~\cite{Bernard:2001rs,bodek08}.
It is reasonable to assume the larger axial mass required by the RFG
as an effective value to incorporate into the calculations nuclear effects
which are not included in the RFG.
A precise knowledge of lepton-nucleus cross
sections, where uncertainties on nuclear effects are reduced as much as
possible, is mandatory and a comparison between the results of different
models can be helpful  to disentangle different physics aspects involved
in the scattering process.

It would be a sound strategy requiring to any nuclear model used to describe
neutrino-nucleus scattering
that they succeed  in the description of available electron scattering
data in similar kinematic region~\cite{Amaro:2004bs}.
At intermediate energy, quasielastic (QE) electron scattering
calculations~\cite{Boffi:1993gs,book}, which are able to
successfully describe a wide number of experimental data, can provide a
useful tool to study neutrino-induced processes.
However, some of these models based on the impulse approximation (IA)
have been shown to be unable to describe
the MiniBooNE data for both CC and NC reactions~\cite{Amaro:2006pr,
Antonov:2007vd,Benhar:2011wy,Ankowski:2012ei}.
This has been viewed as an indication that the reaction can have significant
contributions from effects beyond  the IA.
The contribution of multinucleon excitations to neutrino-nucleus 
scattering~\cite{Martini:2009uj,Martini:2010ex,Martini:2011wp,Martini:2013sha,
Nieves:2011pp,Nieves:2011yp,Nieves201390}
has been found sizable and able to bring the theory in agreement with the
MiniBooNE cross sections without the need to increase the axial mass
$M_A$. On the other hand, a relativistic calculation of 2p2h excitations,
performed for both electron and neutrino scattering
\cite{DePace:2004cr,Amaro:2010sd,Amaro:2011qb,AmaroAntSusa}, has shown that
two body currents give a more modest contribution at MiniBooNE kinematics,
unable to fully account for the data.
Other models invoke an enhancement of the magnetic response rather than a
modification on the axial mass to get agreement with the MiniBooNE
data~\cite{bodek11,Golan:2013jtj}.

A deeper understanding of the reaction dynamics would require a 
careful evaluation of all nuclear effects
and of the relevance of multinucleon emission and of some non-nucleonic
contributions~\cite{PhysRevC.79.034601,Leitner:2010kp,PhysRevC.83.054616,FernandezMartinez2011477,Morfin:2012kn}.
Previous studies have clearly stated the relevance of final state
interactions (FSI) to reproduce
the exclusive $\eep$ cross section within the distorted-wave impulse
approximation (DWIA)~\cite{Boffi:1993gs,book,Udias:1993xy,Meucci:2001qc,
Meucci:2001ja,Meucci:2001ty,Radici:2003zz,Giusti:2011it} and the use of a
complex optical potential (OP). The imaginary part
of the OP produces an absorption that reduces
the cross section and accounts partly for the loss of the incident flux to
the open inelastic channels.
For the case of inclusive scattering, where only the emitted lepton is detected,
%the final nuclear state is not determined, 
all elastic and inelastic channels contribute, and a different treatment of FSI is
required: since all final-state channels are retained, the flux lost in
a channel is redistributed in the other channels and in the sum over all the
channels the total flux must be conserved.

FSI have been considered in relativistic calculations for
the inclusive QE electron- and neutrino-nucleus
scattering under different approaches~\cite{Maieron:2003df,Caballero:2006wi,Caballero:2009sn,Meucci:2003cv,
Meucci:2003uy,Meucci:2004ip,Meucci:2006cx,Meucci:2006ir,Meucci:2008zz,Giusti:2009sy,
Meucci:2009nm,Meucci:2011pi,refId0}.
The simplest one corresponds to the relativistic plane-wave impulse approximation (RPWIA), where FSI
are neglected. In some DWIA calculations FSI effects are incorporated in the final nucleon state by using 
real potentials, either retaining only the real part of the relativistic energy-dependent complex optical potential
(denoted as rROP), or using the same relativistic mean field potential considered in
describing the initial nucleon state (RMF) \cite{Caballero:2005sn,Maieron:2003df}.
Note that the RMF, because of the use of the same strong energy-independent real
potential for both bound and scattering states, fulfills the dispersion
relation~\cite{hori} and maintains the continuity equation.

A different description of FSI involves the use of relativistic Green's function (RGF)
techniques~\cite{Capuzzi:1991qd,Meucci:2003cv,Meucci:2003uy,Capuzzi:2004au,Meucci:2005pk,
Meucci:2009nm,Meucci:2011pi,Giusti:cortona11,esotici2,Meucci:2013gja}.
In the RGF model the components of the nuclear response are written
in terms of the single particle optical model Green's function;
its spectral representation, that is
based on a biorthogonal expansion in terms of a non-Hermitian
OP $\cal H$ and of its Hermitian conjugate $\cal H^{\dagger}$, can be exploited to avoid
the explicit calculation of the single particle Green's function and obtain the
components of the hadron tensor \cite{Meucci:2003uy,Meucci:2003cv}.
Calculations require matrix elements of the same type as the DWIA ones
of the exclusive \eep process in \cite{Meucci:2001qc}, but involve
eigenfunctions of both $\cal H$ and $\cal H^{\dagger}$, where the imaginary
part has an opposite sign and gives in one case a loss and in the
other case a gain of strength. The RGF formalism makes it possible to reconstruct the
flux lost into nonelastic channels in the case of the inclusive response 
starting from the complex OP which describes elastic
nucleon-nucleus scattering data.
Moreover, 
%with the use of the same complex OP, 
a consistent treatment of FSI in the exclusive and in the inclusive scattering is provided, and,
because of the analyticity properties of the OP, the Coulomb sum rule
is fulfilled \cite{hori,Capuzzi:1991qd,Meucci:2003uy}.

A comparison between these different descriptions of FSI has been presented in~\cite{Meucci:2009nm}
for the inclusive QE electron scattering, in \cite{Meucci:2011pi} for
the charged-current quasielastic (CCQE) neutrino scattering,  and
in~\cite{Meucci:2011vd} with the CCQE MiniBooNE data.
The behavior of electron scattering data and their related scaling and
superscaling functions are successfully described by both RMF and RGF models. In the case of neutrinos,
the shape of the experimental CCQE cross sections is well reproduced by both models, although the 
RMF generally underpredicts the CCQE MiniBooNE data, while the RGF can reproduce its magnitude for
some particular choices of the relativistic potential
without the need to increase the standard value of the axial mass.

In this work we extend the comparison between the results of the RGF and
RMF models to NCE scattering.
We note that the RGF is appropriate for an inclusive process where only
the emitted lepton is detected,
whereas  in the NCE scattering the final
lepton is usually not detected and it is the nucleon in the final state what
triggers the event detections. Thus NCE cross sections are usually
semi-inclusive in the hadronic sector,
where events for which at least one nucleon in the final state
is detected are experimentally selected.
The description of semi-inclusive NCE scattering with the RGF approach can
recover important contributions that are not present in the RDWIA, for which the
semi-inclusive cross section is obtained from the sum of all
the integrated single-nucleon knockout channels plus the
absorption produced in each channel by the imaginary part of the optical
potential. This is appropriate for exclusive scattering, but it
neglects some final-state channels which can contribute to the semi-inclusive
reaction. The RGF, however, describes the inclusive process and, as such, may
include channels which are not present in the semi-inclusive NCE measurements. 
From this point of view, the RDWIA can represent a lower limit and the RGF an upper
limit to the semi-inclusive NCE cross sections.
In comparison with the MiniBooNE NCE data, the RDWIA generally underpredicts
the experimental cross section, while the RGF results are in reasonable
agreement with the NCE data \cite{Meucci:2011nc}.

It is not easy to disentangle the role of specific contributions which may be
neglected in the RDWIA or spuriously added in the RGF, in particular
if we consider that both RDWIA and RGF calculations make use of
phenomenological optical potentials, obtained through a fit of elastic
proton-nucleus scattering data.
In order to clarify the content of the enhancement of the RGF cross sections
compared to those of the IA models, a careful evaluation of all nuclear effects
and of the relevance of multinucleon emission and of some non-nucleonic
contributions~\cite {Leitner:2010kp} is required.
The comparison with the results of the RMF model, where only the purely
nucleonic contribution is included, can be helpful for a deeper
understanding of nuclear effects, particularly FSI, which may play
a crucial role in the analysis of upcoming scattering data and of their
influence in studies of neutrino oscillations at intermediate to high energies.

\section{Results \label{results}}

In this section the numerical results of the RGF and RMF models are compared
for NCE neutrino- and antineutrino scattering on $^{12}$C. As a first step, we
have proved that RPWIA cross sections evaluated with two independent computer
programs (developed by the Pavia and Madrid-Sevilla groups)
are almost identical. This gives us enough confidence on the reliability of
both calculations, and it agrees with previous results found
in \cite{Meucci:2009nm} for the inclusive QE
electron scattering and in \cite{Meucci:2011pi} for the CCQE neutrino-nucleus
scattering. Then the comparison between the results corresponding to the RMF
and RGF models
is performed for the  NCE  neutrino and  antineutrino induced cross
sections and also for the ratio between proton- and neutron-knockout cross
sections.
In all the calculations presented in this work the bound nucleon states are
taken as self-consistent Dirac-Hartree solutions derived within a relativistic
mean field approach using a Lagrangian containing $\sigma$, $\omega$, and
$\rho$ mesons~\cite{Serot:1984ey}.

%%%%%%%%%%%%%%%%%%%%%%%%%%%%%%%%%%%%%%%%%%%%%%%%%%%%%%%%%%%%%%%%%%%%%%%%%%%%%%%%
\begin{figure}[t]
    \centering
        \includegraphics[width=.45\textwidth,angle=0]{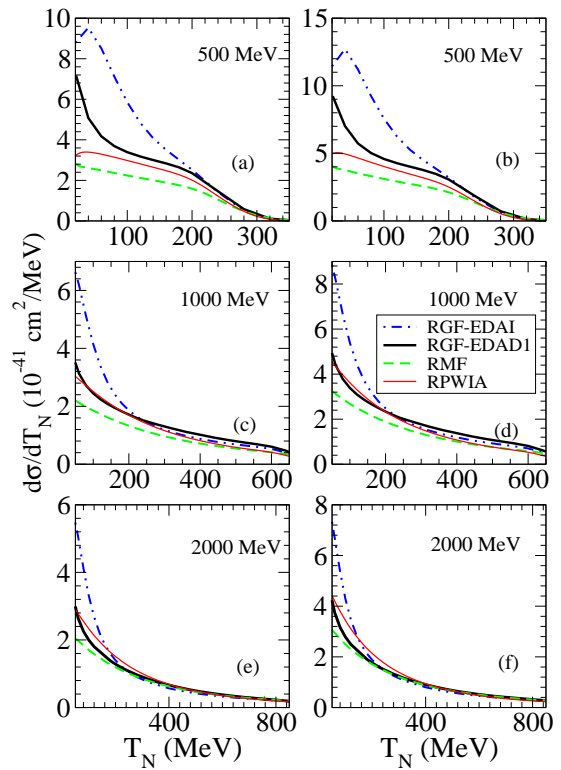}
    \caption{(Color online)
    Differential cross sections of the NCE neutrino scattering on $^{12}$C as a
    function of the kinetic energy of the emitted proton [panels (a), (c), and (e)]
    or neutron [panels (b), (d), and (f)]
    at $\varepsilon_{\nu} = 500, 1000,$ and $2000$ MeV calculated in the
    RPWIA (thin solid lines), RMF (dashed lines), RGF-EDAD1 (thick solid lines), and
    RGF-EDAI (dash-dotted lines).
 The vector and axial-vector strange form factors have been fixed to zero.}
    \label{fig:models_p_et_n_nu}
\end{figure}
%%%%%%%%%%%%%%%%%%%%%%%%%%%%%%%%%%%%%%%%%%%%%%%%%%%%%%%%%%%%%%%%%%%%%%%%%%%%%%%%
\begin{figure}[t]
    \centering
        \includegraphics[width=.45\textwidth,angle=0]{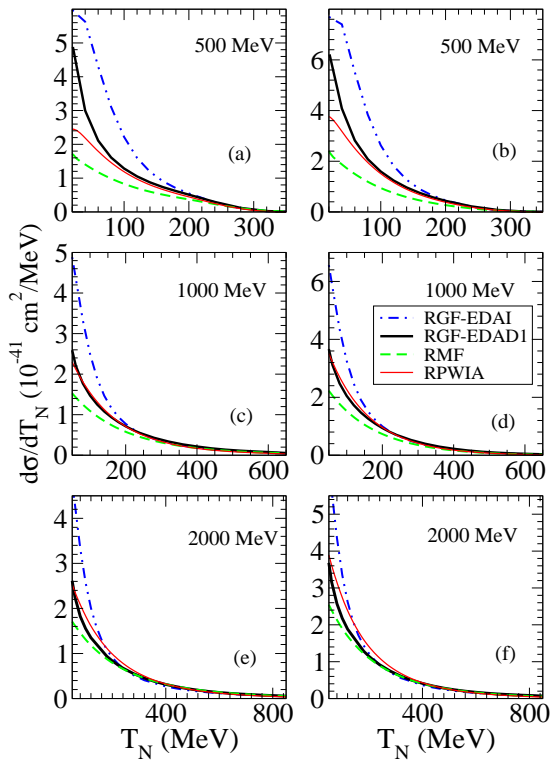}
    \caption{(Color online)
    The same as in Fig. \ref{fig:models_p_et_n_nu}, but for antineutrino
    scattering.
    }
    \label{fig:models_p_et_n_nubar}
\end{figure}
%%%%%%%%%%%%%%%%%%%%%%%%%%%%%%%%%%%%%%%%%%%%%%%%%%%%%%%%%%%%%%%%%%%%%%%%%%%%%%%%

The differential cross sections of the NCE neutrino and antineutrino
scattering, evaluated in the RPWIA, RMF, and RGF, are
presented in Figs. \ref{fig:models_p_et_n_nu} and \ref{fig:models_p_et_n_nubar}
as a function of the kinetic energy of the emitted proton or neutron for three
different (anti)neutrino energies
$\varepsilon_{\nu(\bar \nu)} = 500, 1000,$ and $2000$ MeV.
 The contribution from strange quarks to the vector and axial-vector form
factors has been fixed to zero.
In addition, we note that in all the calculations presented in this work we
have used the standard value of the axial mass $M_A = 1.03$ GeV.
A different value of
$M_A$  would change the cross sections but not the comparison between the
results of the different models.
In the RGF calculations we have used two parametrizations for the
relativistic OP of $^{12}$C: the Energy-Dependent and A-Dependent EDAD1
(where the E represents the energy and the A the atomic number)
and  the Energy-Dependent and A-Independent EDAI
phenomenological OPs  of \cite{Cooper:1993nx}.
The EDAD1 parametrization is a global one, because it is obtained
through a fit to elastic proton-scattering data on a wide range of nuclei and,
as such, it  depends on the atomic number $A$, whereas the EDAI OP  is
constructed   only from elastic proton-$^{12}$C
phenomenology \cite{Cooper:1993nx}. It leads to a better description of the
inclusive QE $^{12}$C$(e, e')$ experimental cross section, as well as to CCQE
and NCE results that are in better agreement with the MiniBooNE
 data within the RGF approach
\citep{Meucci:2009nm,Meucci:2011vd,Meucci:2011nc,Meucci:ant}.

The RMF gives cross sections that are generally $30\%$
lower than the RPWIA ones at small outgoing nucleon kinetic energy $T_N$, but
with a longer tail extending towards larger values of $T_N$, i.e., higher
values of the transferred energy, that is attributable to the strong
energy-independent scalar and vector potentials adopted in the RMF approach.

The RGF cross sections are generally larger than the RPWIA and the RMF ones.
In the RGF the imaginary part of the optical potential redistributes the flux
in all the final-state channels and,  in each channel, the flux lost
towards other channels  is recovered by the flux gained from the other
channels. The larger cross sections in the RGF arise from the translation
to the strength of the overall effects of inelastic channels which are not
included in the other models, such as, for instance, rescattering processes of
the nucleon in its way out of the nucleus,
non-nucleonic $\Delta$
excitations which may arise during nucleon propagation, or also some
multinucleon processes.
 These contributions are not included explicitly in the RGF, but they
all built phenomenologically on the absorptive imaginary part of the OP.
Dispersion relations within the RGF would translate this strength into the
inclusive RGF cross-section.
However,
the RGF is appropriate for an inclusive process where only the emitted
lepton is detected
and can include contributions of channels which are present in an inclusive
but not in a semi-inclusive reaction. From this point of view, the RGF can be considered as an upper limit to the
NCE cross sections.

The comparison between the RGF results obtained with the EDAD1 and EDAI
potentials can give an idea of how  the predictions of the model are
affected by uncertainties in the determination of the phenomenological OP.
The differences depend on the energy and momentum transfer and are essentially
attributable to the different imaginary part of the two
potentials, which accounts for the overall effects of inelastic channels and
is not univocally determined  only from elastic phenomenology.
In contrast, the real term is similar for different parametrizations and gives
similar results.

The NCE experiments can also be used to look for possible strange-quark
contributions in the nucleon.
 The role of strangeness contribution to the electric and magnetic
nucleon form factors has been
recently analyzed for parity-violating elastic electron
scattering~\cite{raulrep}. Specific values for the
electric and magnetic strangeness were provided making use of all available
data at different transferred momenta $Q^2$.
The analysis of $1\sigma$ and $2\sigma$ confidence ellipses showed that zero
electric and magnetic
strangeness were excluded by most of the fits. However, the values of the
strangeness in the electric and magnetic sectors
compatible with the previous study lead to very
minor effects in the separate proton/neutron contribution to the cross section
for neutrino/antineutrino scattering.
Moreover, these {\it \lq\lq small\rq\rq}\ effects tend to cancel being negligible for
the total differential cross sections.
Although this cancellation also works for the axial-vector strangeness, its
relative contribution to the separate
proton/neutron cross section is much larger than the one associated to the
electric/magnetic channels. Therefore,
in this paper we restrict ourselves to the influence of the axial-vector
strangeness and
consider how the NCE antineutrino cross sections change when
the description of the axial-vector form factor of the nucleon is modified.
It is a common prescription to apply the dipole parametrization to the
strange axial form factor
and to use the same value of the axial mass  used for the non-strange
form factor as a cutoff; the strange axial coupling constant  at $Q^2=0$
is $\Delta s$. A measurement of $\nu (\bar{\nu})$-proton elastic scattering at
the Brookhaven National Laboratory at low $Q^2$ suggested
a nonzero value for $\Delta s$ \cite{bnl,gar}.
The MiniBooNE Collaboration used the ratio of proton-to-nucleon NCE
cross sections to extract $\Delta s = 0.08 \pm 0.26$~\cite{miniboonenc}
based on the RFG with $M_A$=1.35 GeV/c$^2$.
The analysis performed in \cite{raulnce} with the RMF model leaded to
$\Delta s = 0.04 \pm 0.28$, while
the COMPASS Collaboration reported a negative
$\Delta s = -0.08 \pm 0.01$ (stat.) $\pm 0.02$ (syst.) as a result of a
measurement of the deuteron spin asymmetry \cite{Alexakhin20078}, in
agreement with the HERMES results \cite{PhysRevD.75.012007}.

%%%%%%%%%%%%%%%%%%%%%%%%%%%%%%%%%%%%%%%%%%%%%%%%%%%%%%%%%%%%%%%%%%%%%%%%%%%%%%%%
\begin{figure}[t]
    \centering
        \includegraphics[width=.27\textwidth,angle=270]{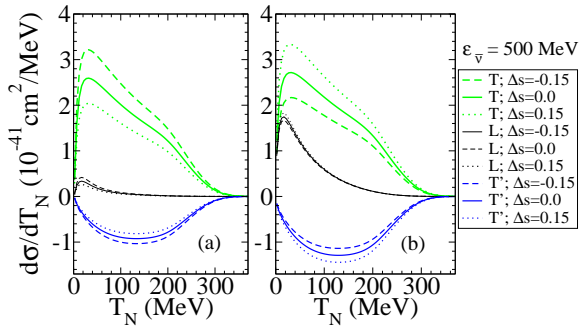}
    \caption{(Color online)
    Separated  longitudinal, $L$ (central set of lines),
transverse { (symmetric)}, $T$ (top set of lines), and
transverse {axial-vector (antisymmetric)}, $T'$ (bottom set of lines), for the
NCE  antineutrino cross
section at $\varepsilon_{\bar \nu} = 500$ MeV as a function of the emitted
proton [panel (a)]
or neutron [panel (b)] kinetic energy. Calculations are performed in the RPWIA.
Solid lines are the results with $\Delta s=0.0$, dashed lines with
$\Delta s=-0.15$, and
dotted lines with $\Delta s=+0.15$.
}    \label{fig:L_T_Tp_antine}
\end{figure}
%%%%%%%%%%%%%%%%%%%%%%%%%%%%%%%%%%%%%%%%%%%%%%%%%%%%%%%%%%%%%%%%%%%%%%%%%%%%%%%%

The (anti)neutrino cross section can be  understood essentially by analyzing
the behaviour of the longitudinal response $L$, the pure vector transverse
response $T$, and the axial-vector transverse response ${T'}$.
In Fig.~\ref{fig:L_T_Tp_antine} the relative importance of {these}
three contributions to the NCE antineutrino differential cross section
is presented for $\varepsilon_{\bar\nu} = 500$ MeV.
For neutrino scattering
the same separation holds but the $T'$ response has opposite sign.
The influence of $\Delta s$
on each response, $L$, $T$, and $T'$, and on separate proton and neutron
events,
is also explored.  In order to avoid complications related to the
description of the FSI and/or to uncertainties due to the particular model,
calculations have been  performed in the RPWIA.
In the case of proton knockout, the transverse
response $T$ is larger by a factor of $\approx 2$ than
the transverse axial-vector response $T'$, and the
longitudinal response
$L$ is very small. In the case of neutron knockout, the $T$ response is still
larger than the $T'$ one but the $L$ contribution is significant.
Note that the longitudinal response is to a large extent insensitive
to strangeness. 

\begin{figure}[t]
    \centering
        \includegraphics[width=.27\textwidth,angle=270]{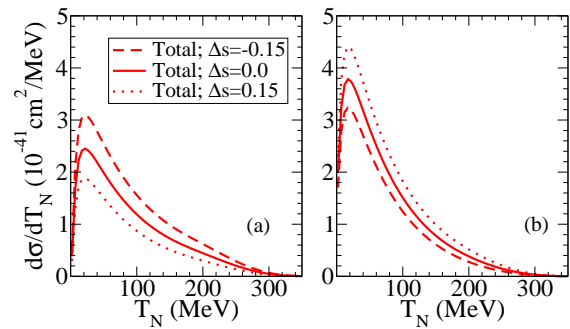}
    \caption{(Color online)
    NCE  antineutrino cross
section at $\varepsilon_{\bar \nu} = 500$ MeV as a function of the emitted
proton [panel (a)]
or neutron [panel (b)] kinetic energy. Calculations are performed in the RPWIA.
Solid lines are the results with $\Delta s=0.0$, dashed lines with
$\Delta s=-0.15$, and
dotted lines with $\Delta s=+0.15$.}
    \label{fig:L_T_Tp_antine3}
\end{figure}

The NCE differential cross sections
are displayed in Fig.~\ref{fig:L_T_Tp_antine3}.
The proton cross section decreases when increasing $\Delta s$, while the neutron
cross section has the opposite behavior. 
Thus, the total proton$+$neutron cross section is almost independent of 
{ $\Delta s$ in the range $-0.15 \div 0.15$.} 
This result is obtained for both neutrino and antineutrino scattering 
and is rather independent of the incident (anti)neutrino energy.

%%%%%%%%%%%%%%%%%%%%%%%%%%%%%%%%%%%%%%%%%%%%%%%%%%%%%%%%%%%%%%%%%%%%%%%%%%%%%%%%
%%%%%%%%%%%%%%%%%%%%%%%%%%%%%%%%%%%%%%%%%%%%%%%%%%%%%%%%%%%%%%%%%%%%%%%%%%%%%%%%
%
%
%%%%%%%%%%%%%%%%%%%%%%%%%%%%%%%%%%%%%%%%%%%%%%%%%%%%%%%%%%%%%%%%%%%%%%%%%%%%%%%%
\begin{figure}[t]
    \centering
        \includegraphics[width=.45\textwidth,angle=0]{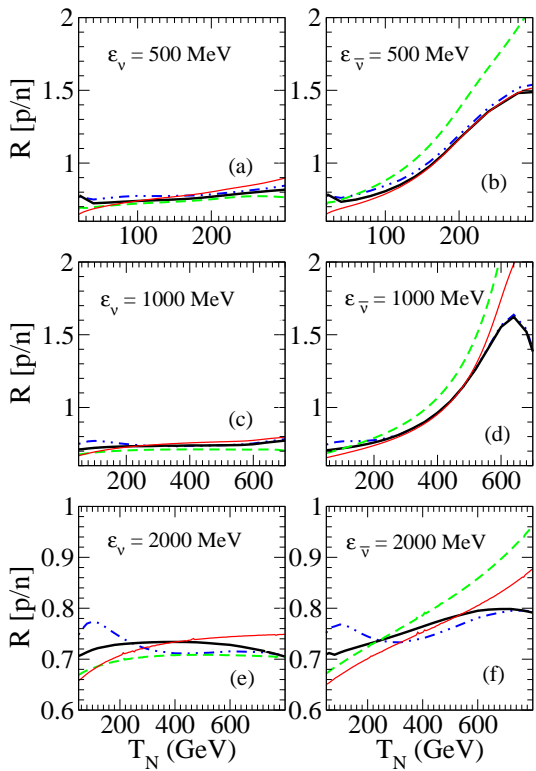}
    \caption{(Color online) Ratio of proton-to-neutron cross sections
    as a
    function of the kinetic energy of the emitted nucleon for
    neutrino [panels (a), (c), and (e)] and
    antineutrino [panels (b), (d), and (f)]. Results of different descriptions
    of FSI are compared.
    Line convention as in Fig.~\ref{fig:models_p_et_n_nu}.
    All the results are obtained with $\Delta s = 0$.}
    \label{fig:ratio_models2}
\end{figure}
%%%%%%%%%%%%%%%%%%%%%%%%%%%%%%%%%%%%%%%%%%%%%%%%%%%%%%%%%%%%%%%%%%%%%%%%%%%%%%%%

A determination of the strangeness contribution to the axial form
factor from measurements of NCE cross sections is not easy. Theoretical
uncertainties on the approximations and on the ingredients of the models
are usually larger than the uncertainty related to the strangeness content of the nucleon.
From the experimental point of view,
precise cross section measurements are not easy due to  difficulties in the
determination of the neutrino flux  related to the nuclear model
dependence.
Therefore, ratios of cross sections have been proposed as alternative and useful
tools to search for strangeness effects.
 The ratio of proton-to-neutron cross sections
was proposed  and discussed
in \cite{gar93,Barbaro:1996vd,Alberico:1997rm,alb02,Lava:2005pb,PhysRevC.74.065501,PhysRevC.76.055501}.
This ratio
 is very sensitive to strange-quark effects because the axial strangeness
 $\Delta s$ interferes
 with the isovector contribution to the axial form factor $g_A \approx 1.27$
 with one sign in
 the numerator and with the opposite
 sign in the denominator.
In Fig. \ref{fig:ratio_models2} we display our results for the $p/n$ ratio for
three different neutrino and antineutrino energies. In the case of ratios of
cross sections
the distortion effects are largely reduced and different models to describe
FSI are expected to produce similar results. To make easier the comparison
between neutrinos and antineutrinos we have chosen the same scale in both
cases. This allows us to visualize clearly the different effects introduced
by the models in both scattering reactions.
In the case of neutrino scattering the $p/n$ ratio is almost constant and the
RPWIA, RMF, and RGF results coincide up to a few percent.
 As observed, in the region of small nucleon kinetic
energy the main difference
in the neutrino case comes from the RGF-EDAI model with a small bump
(for $\varepsilon_\nu=1$ and $2$ GeV) that is not present in the other
approaches.
For larger $T_N$ the ratio stabilizes being the discrepancy between the
different models at most of the order of $\sim 4-5\%$. Finally, the
differences increase at the largest $T_N$ values. Note that in this
region the cross sections are very small and show a significant
sensitivity to FSI and/or the thresholds used. The maximum uncertainty in the
proton/neutron ratio linked to the different models is of the order of
$\sim 15\%$ ($\varepsilon_\nu=500$ MeV) and $\sim 8\%$ ($\varepsilon_\nu=1$
and $2$ GeV).

Larger differences are obtained
in the case of antineutrino scattering, in particular for the RMF model, whose
results are significantly
enhanced with respect to the RGF ones for large values of $T_N$. Contrary to
the case of neutrinos, where the ratio changes very smoothly with $T_N$, for
antineutrinos the slope of the ratio goes up very fast with the nucleon
energy. This reflects the different behavior shown by the proton/neutron
cross sections against $T_N$. At intermediate nucleon energies the uncertainty
between the various models is of the order of $\sim 12-14\%$ getting much
larger discrepancies for
increasing $T_N$-values. However, in this energy region the cross section
becomes significantly lower than its maximum and a very precise measurement is
required to obtain a clear result. It is interesting to point out the
similarity between
the results corresponding to RGF-EDAI, RFG-EDAD1, and RPWIA at
$\varepsilon_{\bar{\nu}} = 500$ and $1000$ MeV.

\begin{figure}[ht]
    \centering
        \includegraphics[width=.45\textwidth,angle=0]{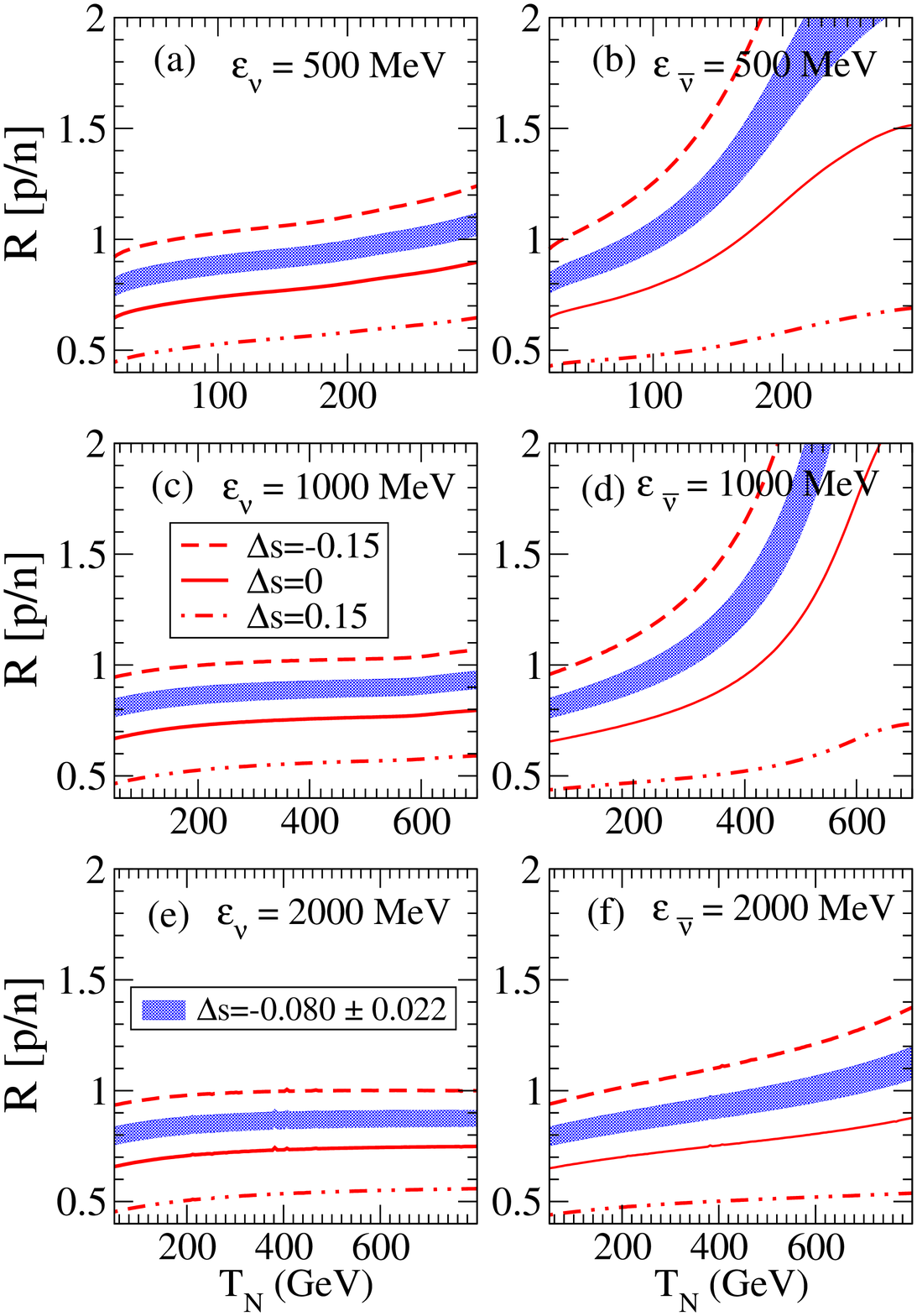}
    \caption{(Color online) Ratio of proton-to-neutron cross sections
    as a
    function of the kinetic energy of the emitted nucleon for
    neutrino [panels (a), (c), and (e)] and
    antineutrino [panels (b), (d), and (f)]. Calculations are performed in the
    RPWIA and with
    different values of $\Delta s$.  The shadowed band refers to results corresponding to
    the COMPASS-HERMES measurement for the axial strangeness.    
    }
    \label{fig:ratio_gas}
\end{figure}

In Fig.~\ref{fig:ratio_gas} the dependence of the  RPWIA $p/n$ ratio
on the
strange-quark contribution is presented. The ratio is enhanced
when calculations are performed with a negative $\Delta s$  and
suppressed when
a positive $\Delta s$ is considered.
In the case of antineutrino scattering the role of strangeness contribution is
particularly
significant when a negative $\Delta s$ is assumed with a
large peak at $T_N \approx 0.7 \varepsilon_{\bar{\nu}}$.
The sensitivity of
the $p/n$ ratio to  $\Delta s $, as well as to the strange-quark contribution
in the vector form factors, was analyzed in \cite{Meucci:2006ir}.
In particular, it was obtained that a moderately large and negative
strangeness contribution to the magnetic moment of the nucleon can cancel
the peak in the $p/n$ ratio. Although a large strangeness contribution to
the vector form factors is not supported by any available experimental
evidence \cite{raulrep}, it would be anyhow intriguing to look for possible
strangeness effects
with a direct measurement of this quantity.
We are aware that a precise measurement of the $p/n$ ratio is a hard
experimental task, but the first measurement of the MiniBooNE
Collaboration \cite{miniboonenc} has proven the validity of this experimental
technique and, hopefully, new data will be available in the future.

In the results of Fig.~\ref{fig:ratio_gas}, the uncertainty in the proton/neutron
ratio associated to the axial strangeness is quite large: in the case of
neutrinos the ratio
changes by a factor 2 when going from positive ($\Delta s=0.15$) to negative
($\Delta s=-0.15$) strangeness. This large range  of variability of  $\Delta s$
is in accordance with $\nu (\bar{\nu})$ Brookhaven data \cite{bnl,gar}
and also with the MiniBooNE results \cite{miniboonenc}, but the
COMPASS measurements suggest a narrower interval for the axial strangeness
\cite{Alexakhin20078} which results in a reduced range of variation of the
proton/neutron ratio. This is represented in Fig.~\ref{fig:ratio_gas} 
by the shadowed band that,
as observed, is of the same order of magnitude as
the uncertainties related to the distortion effects.

This sensitivity to $\Delta s$ gets much larger for antineutrinos,
where the ratio goes up very fast with increasing $T_N$-values.
However, as in the previous case for neutrinos, the range of variation in $R[p/n]$ associated to
the COMPASS measurement is similar to the uncertainty introduced by nuclear model and/or distortion effects.
Although this study is consistent with previous analyses, and it shows the capability
of the ratio $R[p/n]$ as an useful observable in order to get precise
information on the axial-vector strangeness content in the nucleon, the results in
Fig.~\ref{fig:ratio_gas} 
indicate that, owing to the actual precision in the axial strangeness given by
the COMPASS experiment, a deep and careful analysis of the uncertainties linked to 
ingredients of the calculation like nuclear
models or FSI is required.
%\am{I have rewritten the previous sentence}
%
%
%%%%%%%%%%%%%%%%%%%%%%%%%%%%%%%%%%%%%%%%%%%%%%%%%%%%%%%%%%%%%%%%%%%%%%%%%%%%%%%%
%%%%%%%%%%%%%%%%%%%%%%%%%%%%%%%%%%%%%%%%%%%%%%%%%%%%%%%%%%%%%%%%%%%%%%%%%%%%%%%%
%
 \section{Results at MiniBooNE kinematics} \label{minib}

\begin{figure}[ht]
    \centering
        \includegraphics[width=.35\textwidth,angle=270]{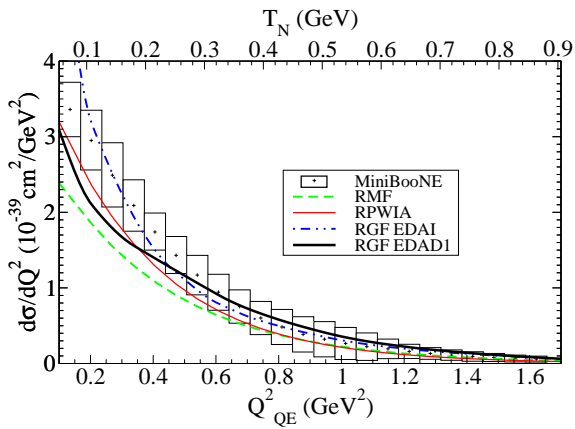}%
     \caption{(Color online)
    NCE flux-averaged $(\nu N \rightarrow \nu N)$  cross section as
    a function of $Q^2$. Line convention as in Fig.~\ref{fig:models_p_et_n_nu}.
    The data are from \cite{miniboonenc}.}
    \label{fig:miniboone_models}
\end{figure}
%%%%%%%
\begin{figure}[ht]
    \centering
     \includegraphics[width=.35\textwidth,angle=270]{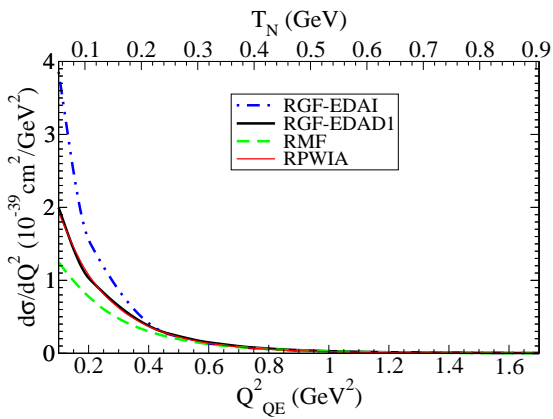}
    \caption{(Color online)
    The same as in Fig. \ref{fig:miniboone_models}, but for
    $(\bar{\nu} N \rightarrow \bar{\nu} N)$ cross section.}
    \label{fig:miniboone_models_nubar}
\end{figure}

The neutrino-nucleus NCE reaction at MiniBooNE
can be considered as scattering of an incident neutrino
or antineutrino with a single nucleon bound in carbon or free in hydrogen.
Each contribution is weighted by an efficiency correction function and
averaged over the experimental (anti)neutrino flux \cite{minibooneflux}.
Different relativistic descriptions of FSI were  presented and compared with
the NCE MiniBooNE data in \cite{Meucci:2011nc,raulnce}.
In  Fig.~\ref{fig:miniboone_models} we present our RMF and RGF cross sections
for NCE $(\nu N \rightarrow \nu N)$ scattering and compare them with the
experimental data,
where the variable $Q^2_{QE}=2m_NT$ is defined assuming that the
target nucleon is at rest, $m_N$ being the nucleon mass and $T$ the total
kinetic energy of the outgoing nucleons.
The RMF result has a too soft $Q^2$ behavior to
reproduce the
experimental data at small $Q^2$, while the RGF produces larger cross sections
in better agreement with the data.
The difference between
the RGF results calculated with the two optical potentials is
significant, particularly for
small $T_N$ ($Q^2_{QE}$) values. This is consistent with the large
discrepancies shown by the cross sections evaluated at fixed
neutrino/antineutrino energies
 (see Fig.~\ref{fig:models_p_et_n_nu}).
The RGF-EDAI cross section is in accordance
with the shape and the magnitude of the data. On the contrary, the RGF-EDAD1
lies below
the data at the smallest values of $Q^2$ considered in the figure.
The RMF approach leads to the lowest
cross section for low-to-intermediate values of the transferred four-momentum.
Only for $Q^2_{QE}\geq 0.9$ GeV$^2$
the RMF tail is higher than the RPWIA result, but still being below the two
RGF models. However, in
this kinematical regime all the models are able to reproduce the data within
the error bars.

The MiniBooNE Collaboration has collected also an extensive data set
of  neutral current antineutrino events whose
analysis is currently ongoing and some preliminaries results are
available \cite{nubarthesis,grangenuint}.
In Fig.~\ref{fig:miniboone_models_nubar} we
show our predictions for the NCE MiniBooNE
$(\bar{\nu} N \rightarrow \bar{\nu} N)$ cross section.
In these calculations we use the set  of efficiency coefficients given in
\cite{miniboonenc} for neutrino scattering.
The selection for the antineutrino NCE sample is slightly different from the
neutrino sample, and therefore the efficiencies are similar only as a first
approximation, since they are expected to be a little bit different.
However, even if it is not rigorous, the use of neutrino efficiencies for
the antineutrinos is approximately  correct.
Similarly to the neutrino case, the RMF gives cross sections that are
lower than the RPWIA ones whereas the RGF produces larger cross sections.
This is consistent with the results shown in
Fig.~\ref{fig:models_p_et_n_nubar} for fixed antineutrino
energies, where a significant discrepancy between the cross sections obtained
with the various models is observed, being the smallest contribution for the
RMF and the largest one for RGF-EDAI.
Furthermore, the RGF with the EDAD1 optical potential gives results which are
very similar to the RPWIA calculation.  The predictions of these two models, RPWIA
and RGF-EDAD1, agree very well with the preliminary antineutrino
NCE MiniBooNE data~\cite{nubarthesis,grangenuint} 

%%%%%%%
\begin{figure}[ht]
         \includegraphics[width=.34\textwidth,angle=0]{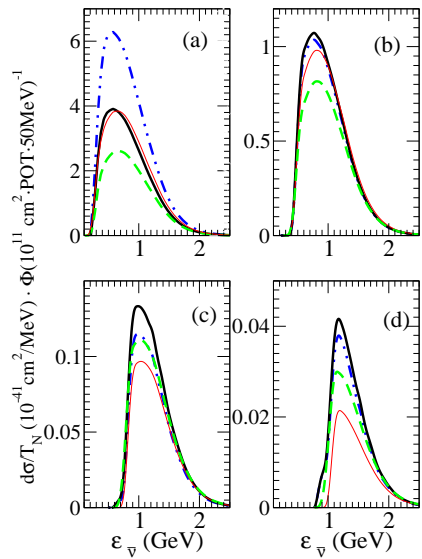}
     \caption{(Color online)
     Product of the proton$+$neutron NCE antineutrino cross section and the antineutrino
     MiniBooNE flux \cite{minibooneflux} as a function of the antineutrino energy
     $\varepsilon_{\bar\nu}$
     at four fixed values of the outgoing nucleon kinetic energy:
     $T_N = 108$ [panel (a)], $T_N = 252$ [panel (b)], $540$ [panel (c)], and
     $756$ MeV [panel (d)].
     Line convention as in Fig.~\ref{fig:models_p_et_n_nu}.
     \label{fig:tn_fijado}}
   \end{figure}

The curves displayed in 
Figs.~\ref{fig:miniboone_models} and \ref{fig:miniboone_models_nubar}
involve a convolution over the experimental (anti)neutrino flux.
In order to better understand these results,
in Fig.~\ref{fig:tn_fijado} we present the proton$+$neutron NCE antineutrino
cross section multiplied by the antineutrino MiniBooNE
flux of \cite{minibooneflux} as a function of the antineutrino energy
for four different values of the kinetic energy of the emitted nucleon.
The calculations required for the analysis in Fig.~\ref{fig:tn_fijado}
consider the entire energy range which is
relevant for the MiniBooNE  flux.
It has been pointed out in~\cite{Benhar:2010nx,Benhar:2011wy} that the
flux-average procedure
introduces additional uncertainties  and, therefore,
the MiniBooNE cross sections can include contributions from different
kinematic regions, where
other reaction mechanisms than one-nucleon knockout can be dominant.
Part of these contributions, which are not included in usual calculations
based on the IA, can be recovered in the RGF by the imaginary part of
the phenomenological OP.
The RMF gives cross sections that are lower than the RPWIA ones at
$T_N = 108$ and $252$ MeV, but larger at higher values of $T_N $.
As already mentioned, this effect is due to the strong energy-independent
potential adopted in the RMF model.
The larger cross section in the RGF can be ascribed to the contribution of
reaction channels which are not included in other models based on the IA.

%\am{I have rewritten the following paragraph }
The MiniBooNE Collaboration has also reported the $(\nu p \to \nu p)/(\nu N \to \nu N)$ 
ratio \cite{miniboonenc}. 
The denominator of this ratio includes events with standard NCE selection cuts 
but with the energy cut replaced with 
$350~\text{MeV}<T_N <800~\text{MeV}$, and an additional \lq\lq proton/muon\rq\rq\  
cut in order to reduce muonlike backgrounds that dominate the 
high visible energy region. In the numerator are events from so called 
\lq\lq NCE proton-enriched event sample\rq\rq\ where two additional cuts are applied to 
suppress neutron NCE events. 
The MonteCarlo simulation shows that only 10\% neutron NCE events give 
contribution to the $\nu p \to \nu p$ sample. 
More details on the folding procedure to calculate this ratio are given in
Appendix B of~\cite{Perevalov2009}.

\begin{figure}[t]
\includegraphics[width=.75\columnwidth,angle=270]{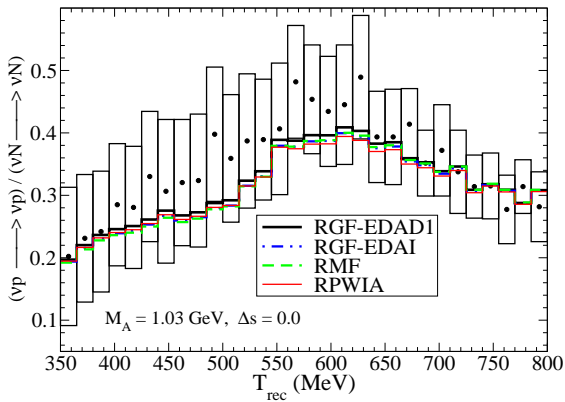}
\caption{(Color online) Ratio $(\nu p \to \nu p)/(\nu N \to \nu N)$ as a
function of the reconstructed energy computed within 
RGF, RMF and RPWIA models. Line convention as in Fig.~\ref{fig:models_p_et_n_nu}. 
The data are from  \cite{miniboonenc}.\label{fig:ratio_models}}
\end{figure}

In Fig.~\ref{fig:ratio_models} we present our results for the 
$(\nu p \to \nu p)/(\nu N \to \nu N)$ ratio with RGF, RMF and RPWIA models as a function 
of reconstructed energy $T_{{rec}}$. In our calculations the axial 
strangeness $\Delta s$ has been fixed to $0$.
All the models  give very close results which are in agreement with experimental data 
within error bars; this is in accordance with the fact that in this kinematical regime with
$T_N>350$~MeV all the models are able to reproduce the cross sections 
data.

\section{Conclusions}

This work extends { previous comparative studies to include
the analysis of neutral-current (NC) neutrino-nucleus scattering reactions. }
In previous works we applied our models
to inclusive electron
and charged-current (CC) neutrino scattering, providing also a comparison with
data measured by the MiniBooNE collaboration.
Our main objective in this paper is to examine how capable our theoretical
models are to explain the recent data on NC
reactions measured by MiniBooNE. In both cases, CC and NC processes, the
kinematics involved implies the use of fully relativistic models.
This is the case of the Relativistic Mean Field (RMF) and the Relativistic
Green's Function (RGF) approaches considered in this work. Not only relativistic
kinematics is considered, but also nuclear dynamics and current operators are
described within a relativistic formalism. Moreover, final state interactions
(FSI), an essential ingredient in the reaction mechanism, are also taken
into account by introducing relativistic potentials in the final-state and
solving the Dirac equation. Whereas in the RMF case the potential consists
of real strong energy-independent scalar and vector terms (the same used for
the bound nucleon states), the RGF makes use of phenomenological
energy-dependent complex optical potentials. In this work results are shown
for two choices of the optical potential: EDAI and EDAD1.

We have compared the predictions for the differential cross sections and the
proton/neutron (p/n) ratio. The former shows an important dependence with the
model, particularly at small values of the outgoing nucleon kinetic energy.
The RMF provides the lowest result while the RGF gets much more strength,
although a significant dependence on the potentials considered is
also seen for the RGF case. This general result applies to both neutrino and
antineutrino reactions, and occurs for very different values of the lepton
($\nu_\mu/\overline{\nu}_\mu$) energy. This explains the significant
differences observed for the NC flux-averaged cross sections which are also
compared with MiniBooNE data. From our analysis we conclude that the largest
contribution corresponding to RGF-EDAI is in accordance with data for
neutrinos, whereas the other models, in particular the RMF, lie clearly
below data at small nucleon kinetic energies ($T_N$). On the contrary, all
models reproduce the behavior of data at larger $T_N$-values. However, we
have to keep in mind
the large data error bands in this kinematical regime.

In addition to the uncertainties associated to nuclear model and/or FSI
descriptions, that are particularly relevant for the cross sections, another
ingredient to be carefully considered is the role of strangeness in the
nucleon.
{ While
strangeness in the electric and magnetic sectors
leads to very minor effects, almost negligible for the total cross section,
the dependence upon the axial-vector strangeness is much more
important.}
This is particularly true in the case of the separate proton/neutron
contributions to the cross sections.
The role of the axial strangeness is opposite in protons and neutrons,
and tends to be cancelled in the total cross section.
This justifies the use of total cross sections to analyze nuclear models and
FSI dependences, being almost independent of $\Delta s$ (axial strangeness).
Moreover, it also  justifies the use of the p/n ratio as a useful
observable to get
information on the axial strangeness.

In this work we have analyzed in detail the proton/neutron ratio comparing the
predictions given by the RMF and RGF models. We have proved that the ratio only
presents a weak dependence on the model, in particular, in the case of
neutrinos: the uncertainty is on average of the order of $\sim 4-5\%$.
This discrepancy gets significantly higher for antineutrinos at increasing
values of the nucleon energy. In any
case, these differences are usually  smaller than the ones ascribed to the use
of different axial strangeness content in the nucleon. In this case the p/n
ratio can change by more than a factor 2 { when the variation in $\Delta s$ is in accordance
with the Brookhaven and MiniBooNE data. However, the highly precise measurements given by COMPASS lead to
an uncertainty in $R[p/n]$ similar to the one ascribed to distortion/nuclear model effects.}

Summarizing, we have applied two different relativistic models that
incorporate final state interactions to the study of NCE neutrino
(antineutrino)-nucleus scattering processes. We have presented a detailed
analysis of the differential cross sections (with the separate proton and
neutron contributions) and the p/n ratio. We have compared our predictions
with the recent experimental data taken by MiniBooNE collaboration for
neutrinos, and given predictions for antineutrinos which can be also compared
with data when available. We have proved the significant differences
introduced by the various models that may indicate important effects
ascribed to correlation and Meson Exchange Currents, not yet
incorporated in the models.
Although the comparison between RMF and RGF may help us in disentangling
different aspects involved in the physics of the problem, we should be
cautious in getting final conclusions before other ingredients beyond the
impulse approximation can be implemented in more refined calculations, and
their contributions are carefully examined.

%%%%%%%%%%%%
\begin{acknowledgments}

This work was partially supported by the Italian MIUR through the PRIN 2009 research project,
by the Istituto Nazionale di Fisica Nucleare under Contract MB31,
by Spanish DGI and FEDER funds (FIS2011-28738-C02-01, FPA2010-17142), by the Junta
de Andalucia, by the Spanish Consolider-Ingenio 2000 program CPAN (CSD2007-00042), by the Campus of Excellence
International (CEI) of Moncloa project (Madrid) and Andalucia Tech, by the INFN-MICINN collaboration agreement (AIC-D-2011-0704),
as well as by the Bulgarian National Science Fund under contracts No. DO-02-285 and DID-02/16-17.12.2009.
M.V.I. is grateful for the warm hospitality given by the UCM and for financial
support during his stay there from the SiNuRSE action within the ENSAR european project.
R.G.J. acknowledges support from the Ministerio de Educaci{\'o}n (Spain).
\end{acknowledgments}

%%%%%%%%%%
\bibliography{rif-pv-sevilla}

\end{document}